\documentclass[11pt,letterpaper,notoc,nohyper]{JHEP3}
\input epsf.tex
\usepackage{graphics}
\usepackage{amsmath}
\usepackage{cite}
\usepackage{colordvi}

\DeclareMathOperator{\tr}{tr}
\DeclareMathOperator{\diag}{diag}
\newcommand\openone{\leavevmode\hbox{\small1\normalsize\kern-.33em1}}
\providecommand{\abs}[1]{\lvert#1\rvert}
\def\CL{{\cal L}}

\def\CO{{\cal O}}
\def\CL{{\cal L}}

\def\one{{\bf 1}}

\def\be{\begin{equation}}
\def\ee{\end{equation}}
\def\bea{\begin{eqnarray}}
\def\eea{\end{eqnarray}}

\newcommand{\gsim}{ \mathop{}_{\textstyle \sim}^{\textstyle >} } 
\newcommand{\lsim}{ \mathop{}_{\textstyle \sim}^{\textstyle <} } 

\newcommand{\vev}[1]{ \left\langle {#1} \right\rangle }

\newcommand{\Tr}{ {\rm Tr} }

\newcommand\half{\frac{1}{2}}
\newcommand\beq{\begin{eqnarray}}
\newcommand\eeq{\end{eqnarray}}

\usepackage{epsfig}
\usepackage{latexsym}
\title{The Intermediate Higgs}
\author{Emanuel Katz \\  
            Department of Physics, Boston University, Boston, MA 02215,
	    \\Theory Group, Stanford Linear Accelerator Center, Menlo Park, CA 94025, 
                        \\ email:  \email{amikatz@slac.stanford.edu, amikatz@bu.edu}}
\author{Ann E. Nelson \\ Department of Physics, 
            Box 1560, 
            University of Washington, 
            Seattle, WA 98195-1560
 , \\ email: \email{anelson@phys.washington.edu}}
\author{Devin G. E. Walker\\  Jefferson Laboratory of Physics, 
	    Harvard University, 
            Cambridge, MA 02138 
  \\ email: \email{walker@lamb.physics.harvard.edu}} 
  
\preprint{BUHEP-05-06,\\HUTP-05/A0018,\\SLAC-PUB-11111,\\UW/PT-05/06.} 
  
\abstract{
Two paradigms for the origin  of electroweak superconductivity are a weakly coupled scalar condensate,  and a strongly coupled fermion condensate. The former  suffers from a finetuning problem unless there are  cancelations to radiative corrections, while the latter presents potential discrepancies with precision electroweak physics.  Here we present a framework for  electroweak symmetry breaking  which interpolates  between these two paradigms, and mitigates their faults.  As in Little Higgs theories,  the Higgs is a pseudo-Nambu Goldstone boson, potentially composite. The cutoff sensitivity of the one loop top quark contribution to the effective potential is canceled by contributions from additional  vector-like quarks, and the cutoff can naturally be higher than in the minimal Standard Model. Unlike the Little Higgs models,  the cutoff sensitivity from one loop gauge contributions is not canceled. However, such gauge contributions are naturally small as long as the cutoff is below 6 TeV.  Precision electroweak corrections are suppressed relative to those of Technicolor or generic  Little Higgs theories.  In some versions of the intermediate scenario, the Higgs mass  is computable in terms of  the masses of these additional fermions and the Nambu-Goldstone Boson decay constant. In addition to the Higgs, new scalar and pseudoscalar particles are typically present at the weak scale.}

\begin{document}

\section{Introduction}  

The origin of electroweak superconductivity is the central issue in high energy particle physics.  The minimal Standard Model accounts well for this phenomena, and is in agreement with current particle physics experiments.  However, because scalar masses are quadratically sensitive to short distance physics, the required value of the Higgs expectation value is produced by finetuning the mass squared parameter in the Higgs potential to a precision of order (1~TeV$/\Lambda$)$^2$, where $\Lambda$ is the scale of new physics.  Here  a TeV is the ``'t Hooft scale" 
 \cite{thooft} at which new physics must come in if finetuning is to be avoided. The ``little hierarchy problem'' is that the success of the Standard Model in predicting the value of precision electroweak observables is only guaranteed if $\Lambda$ is at least several TeV. Thus either finetuning of physical parameters  is acceptable in nature, or we must seek some clever sort of new physics which reduces the sensitivity of the Higgs potential to short distance physics but does not produce large precision electroweak corrections. The canonical  example of such a clever theory is the Minimal  Supersymmetric Standard Model (MSSM).  Since there is no unambiguous evidence for supersymmetry, and since weak scale supersymmetry entails many new puzzles, it is worth searching for  alternatives.

 In this paper we address the little hierarchy problem in a non-supersymmetric model which has minimal impact on precision electroweak measurements. We model the Higgs as a pseudo-Nambu-Goldstone boson (PNGB), as in Composite Higgs  (CH) 
 \cite{Kaplan:1983sm,Georgi:1984af,Georgi:1984ef,Dugan:1984hq} and  Little Higgs  (LH) models \cite{hep-ph/0105239,Arkani-Hamed:2002qy,Arkani-Hamed:2002qx,Low:2002ws,Kaplan:2003uc,Chang:2003un,Skiba:2003yf,Chang:2003zn}. CH models interpolate between strongly coupled and weakly coupled electroweak symmetry breaking theories, and  resemble the minimal Standard Model in a decoupling limit where the PNGB decay constant $f$ is large compared with the Higgs vev $v$. In the  limit where $f\approx v$, the symmetry breaking sector is strongly coupled at the weak scale, as in technicolor\cite{Weinberg:1975gm,Susskind:1978ms,Weinberg:1979bn}. The innovation of the LH models is a natural mechanism for $v^2$ to be less than $f^2$ by a loop factor of order $1/(4\pi)^2$. The LH mechanism  requires new gauge bosons, scalars and fermions at or below  the scale $f$, which cancel one loop quadratically divergent corrections to the Higgs mass squared.  These new particles can give tree level corrections to precision electroweak observables, which are of about the same size as one loop corrections in the minimal Standard Model, and which produce potential experimental discrepancies. The largest such corrections come from the new bosons \cite{hep-ph/9903476,hep-ph/9903451,hep-ph/0211124,hep-ph/0211218,hep-ph/0301040,hep-ph/0303236,hep-ph/0305275,hep-ph/0311038,hep-ph/0311095,hep-ph/0312053,hep-ph/0502096}, although there are natural mechanisms for suppressing these corrections \cite{hep-ph/0307339,hep-ph/0308199,hep-ph/0311095,hep-ph/0409025}.

In LH models, the motivation for   the new bosons is much less compelling  than the motivation for the new fermions, and the price paid for their introduction is rather high.  The new bosons have couplings to the Higgs which naturally cancel the one loop quadratic divergences from the gauge sector. However, the largest radiative contributions to the Higgs potential arise not from the gauge bosons but from the top quark, due to both its large Yukawa coupling and the multiplicity of states. Numerically, the top contribution dominates the gauge  contribution  to the quadratic sensitivity by a factor of 
about 10.  With the top contribution uncanceled, the cutoff of the theory would have to be about 2 TeV to maintain naturalness. Canceling the top contribution allows the cutoff to be raised to about 6 TeV, which is high enough to suppress precision electroweak corrections from cutoff scale physics  to experimentally acceptable levels\footnote{With a cutoff as low as 6 TeV, it is still necessary to assume that cutoff scale physics does not induce large flavor, CP or custodial $SU(2)_c$ violation.}.  Furthermore, the constraints from precision electroweak corrections on new bosons are much more severe than the constraints on new fermions, due to the fact that new bosons can  give tree level contributions to many electroweak observables. Thus eliminating the new gauge bosons eliminates a potentially dominant source of precision electroweak corrections.  In addition, model building for a UV completion is much simpler without an extended electroweak gauge symmetry.
We  therefore propose an ``intermediate'' Higgs (IH) model, where the one loop cutoff sensitivity from the top quark contribution to the potential is canceled  by heavy quark states, but the gauge contribution is left uncanceled.   In the IH model, the decay constant $f$ can be as large as 5 times  $v$  without  finetuning.   One   then has an effective theory with a cutoff below 6 TeV, which is not finely tuned, as long as new quark states are lighter than 2 TeV.  Relative to technicolor-like models, precision electroweak corrections are suppressed by a factor of order $(v/f)^2$. Since the precision electroweak corrections in technicolor theories are typically only  too big by a factor of a few \cite{Golden:1990ig,Holdom:1990tc,Peskin:1990zt,Peskin:1991sw,Appelquist:1991is},   such a suppression factor is  sufficient\footnote{In certain extended technicolor theories with very large flavor symmetries and numerous PNGBs, the precision electroweak corrections are much larger.}.

We work out some of the details of a few simple IH models. An economical model, with only one additional weak scale boson,  has  the symmetry breaking pattern $SU(4)/Sp(4)$. In this model there is a new pseudoscalar with a mass of order the weak scale, which nontrivially impacts Higgs search strategies.   Another interesting model  has symmetry breaking pattern $SU(5)/SO(5)$. This model contains 10 additional scalar states at the weak scale, which are  electroweak triplets and singlets.  A softly broken custodial $SU(2)_c$ symmetry constrains the potential and protects the $T$ (or $\rho$) parameter against large corrections.   Only the Higgs doublet gets a vev, and the  Higgs mass is calculable in terms of the masses of new quark states.    We also feature a two Higgs doublet model based on the coset space $SU(6)/Sp(6)$, in which the  $T$ parameter is similarly protected.  In addition to the new Higgs doublets, there are additional charged and neutral scalar states. 

This paper is organized as follows.  In section~\ref{sec: IH}, we discuss the blueprint for constructing this class of models.  Our three examples  are presented in section~\ref{sec:addIH}.  We devote section~\ref{sec: coll} to mention some implications of the models on various collider signatures.  Section~\ref{sec: exp}  reviews the relevant experimental constraints from, e.g., precision electroweak measurements and flavor changing neutral currents. Section~\ref{sec: UV} discusses ways to UV complete these types of models.  A summary is offered in section~\ref{sec: con}.

\section{The Intermediate Higgs}
\label{sec: IH}  

\subsection{The Blueprint}
 \label{sec: blue}
We want to realize the Higgs as a pseudo-Nambu Goldstone boson whose low energy interactions are described by an effective chiral lagrangian.  We begin by embedding the electroweak gauge group $SU(2)_L \otimes U(1)_Y$ into an approximate global symmetry group $G$, which is spontaneously broken to $H$.  $H$ is required to contain the diagonal, approximate ``custodial" $SU(2)_c$.  The space of nearly degenerate candidate vacua is given by the coset $G/H$. The various small terms which explicitly break the symmetries  will determine the ``vacuum alignment," that is, which of the  candidate vacua is the exact minimum of the potential. Electroweak superconductivity is thus an issue of vacuum alignment. When H aligns so that the $SU(2)_L \otimes U(1)_Y$ gauge group is not Higgsed, the PNGBs which parameterize the coset space $G/H$  should contain at least one field with the electroweak quantum numbers of the Standard Model Higgs doublet.  The vev of this doublet  parameterizes the orientation of the ground state with respect to $SU(2)_L \otimes U(1)_Y$.  In the limit where this vev is small compared with the PNGB decay constant $f$, the weak scale physics is weakly coupled and similar to a minimal extension of
the Standard Model with  an extended scalar sector.

A key requirement is that a remnant approximate custodial symmetry $SU(2)_c$, which is the diagonal subgroup of $SU(2)_L\otimes SU(2)_R$, is not broken by the vacuum alignment.  The custodial $SU(2)_c$ suppresses corrections to the experimentally verified relation $m_z = m_w \cos{\theta}$.  The Standard Model violates custodial symmetry by Yukawa couplings and the hypercharge coupling.  Non-standard one loop radiative corrections to the $\rho=1$ relation are  tolerable if suppressed by the product of a loop factor and $(v/f)^2$, or if suppressed by a loop factor containing a sufficiently small coupling. Models in which corrections to $\rho=1$ occur at tree level and are   suppressed only by $\CO(v/f)^2$ corrections are unacceptable in  IH models, since IH models are finely tuned unless $v/f\gsim 0.2$.

The Higgs transforms nonlinearly under $G$, and so its potential arises entirely from small terms which  break  $G$ explicitly, such as the $SU(2)_L \times U(1)_Y$ electroweak gauge interactions, and the Yukawa interactions. Symmetry breaking interactions which are important at short distance are called ``hard", and  renormalization of the quantum corrections proportional to these couplings will require the introduction of new, incalculable parameters.  The small symmetry breaking terms, such as the Yukawa couplings of the light quarks and leptons and the weak gauge couplings, are ``hard". The associated new parameters are not calculable within the effective theory, but being proportional to small couplings, they are naturally  small.

Collective symmetry breaking\cite{Georgi:1975tz,hep-ph/0105239} is a mechanism for ``soft" (only important at low energy) breaking of a nonlinearly realized symmetry.
The basic idea of the LH application of collective symmetry breaking is that two or more couplings are required to be nonzero in order to break all the symmetry which  protects the Higgs mass against radiative corrections\cite{hep-ph/0105239,Arkani-Hamed:2002qx}. Then quadratically divergent contributions to the Higgs mass require contributions from more than one coupling, and are not present at one loop.  To apply the collective symmetry breaking mechanism to the top mass,  we couple vector-like charge 2/3 quarks to the Higgs in a manner which respects a symmetry under which the Higgs transforms nonlinearly. The top Yukawa coupling  arises from  additional terms  which mix  the top with these additional  fermions\footnote{This mass generation mechanism is similar to Frogatt-Nielson models of flavor \cite{Froggatt:1979nt} and the top see-saw mechanism \cite{Chivukula:1998wd}}.  The associated symmetry breaking  induces calculable corrections to the effective potential for the PNGBs which are independent of short distance dynamics.   The main experimental consequence is the presence of at least one new charge 2/3 quark with vector-like electroweak quantum numbers. Corrections to the quadratic term in the Higgs potential are proportional to the mass squared of this quark, and so in a natural theory at least one new quark should be lighter than  about 2 TeV.

\section{Sample Intermediate Higgs Models}
\label{sec:addIH}
We give here several examples of intermediate Higgs models, which   illustrate distinct possibilities for the spectrum of weak scale scalars.  The common features include new charge 2/3 quarks, new scalars, and custodial $SU(2)_c$ constraints on the scalar potential. Custodial $SU(2)_c$, and constraints from the $\rho$ parameter play a significant role in determining an acceptable model. In particular, our second and third  examples demonstrates why it is not sufficient  to identify an approximate $SU(2)_R$ symmetry. Even if  the global symmetry $G$ contains an approximate $SU(2)_R$ symmetry,  it is necessary that a diagonal $SU(2)_c$ subgroup of the $SU(2)_R$ and the electroweak $SU(2)_L$ is unbroken in the ground state.   In the simplest model, where the PNGBs parameterize the coset space $SU(4)/SP(4)$, the preservation of $SU(2)_c$ is automatic. In models with additional   electroweak doublets and/or triplets, generally there exist alignments which preserve the electromagnetic U(1) while breaking the $SU(2)_c$. Since $SU(2)_c$ is always  explicitly broken by the sector which generates the top Yukawa coupling, and since vacuum alignment is affected by the Yukawa sector contribution, models with additional weak scale doublets or triplets are very constrained. Our second example, based on $SU(5)/SO(5)$, illustrates how a model with additional weak scale triplets can avoid tree level corrections to the $\rho$ parameter. Our final example, based on $SU(6)/SP(6)$, contains two Higgs doublets. Such IH models generically preserve $SU(2)_c$ only for a particular ratio of the two vevs. However our $SU(6)/SP(6)$ model actually embeds weak hypercharge into an approximate  $(SU(2)_R)^2$ subgroup of $SU(6)$, ensuring that  any ratio of the two Higgs vevs preserves a remnant  $SU(2)_c$, and there are no tree level corrections to $\rho$.

\subsection{The $SU(4)/Sp(4)$ model}
\label{sec: SUSP}

We describe in this section a simple model based on coset space $G/H= SU(4)/Sp(4)$.  Under  $SU(2)_L \otimes SU(2)_R$, the five PNGBs transform as
\be
 (1,1) \oplus (2,2)
\ee
$U(1)_Y$ is a subgroup of $SU(2)_R$. 
Under $SU(2)_L \otimes U(1)_Y$, the  PNGBs transform as
\be
 1_0 \oplus 2_{\pm1/2}
\ee
giving a real  electroweak singlet and a standard Higgs  doublet. The vev of this doublet preserves the custodial $SU(2)_c$ symmetry -- the diagonal subgroup of  $SU(2)_L \otimes SU(2)_R$.  This guarantees that $\rho=1$ at tree level.
We may use a nonlinear sigma model to describe the low energy effective theory.  We can describe  the vacuum orientation   via an antisymmetric unitary matrix $\Sigma$, which transforms under $SU(4)$ as 
\be
\Sigma \to V \Sigma V^T
\ee
where $V$ is an $SU(4)$ matrix.   It is convenient to specify a background field $\Sigma_0$,
\be
\Sigma_0 = 
 \begin{pmatrix} 
    i\sigma_2 & \quad \\ \quad  & i\sigma_2 \\ 
      \end{pmatrix}.
\ee
which is invariant under the $SP(4)$ subgroup containing $SU(2)_L \otimes SU(2)_R$. In this background, the unbroken $SP(4)$ generators $T$ satisfy
\be 
T \Sigma_0+ \Sigma_0 T^T=0
\label{eq: SUSPunbroken}
\ee
and the broken generators $X$ satisfy
\be 
X\Sigma_0-\Sigma_0 X^T=0.
\label{eq: SUSPbroken}
\ee
The Nambu-Goldstone bosons are fluctuations about this background in the direction of the broken generators, $\Pi \equiv \pi^a X^a$.  They are parameterized as 
\be
  \Sigma(x) = e^{i \Pi/f} \Sigma_0 e^{i \Pi^T/f} =  e^{2 i \Pi/f}
  \Sigma_0 .
\ee
  This non-linear sigma model contains momentum dependent interactions suppressed by the decay constant $f$.  The cutoff, $\Lambda$, may be as as high as $4\pi f$, where the model becomes strongly coupled.  The weak gauge group  $SU(2)_L \times U(1)_Y \subset SU(4)$ is generated by
  \begin{align}
  Q^a =
  \begin{pmatrix}
    \sigma^a/2 &\quad \\ \quad&\quad
  \end{pmatrix}, & &
  Y =  \diag(0,0, 1,-1)/2.
\label{SUSPgauge}
\end{align}
The overall global custodial symmetry is  generated by
\be
 R^a =
  \begin{pmatrix}
    \quad&\quad \\ \quad& \sigma^{a}/2
  \end{pmatrix}.
\label{SUSPcustodial}
\ee
These generators remain unbroken in the reference vacuum $\Sigma=\Sigma_0$.   The PNGB matrix may be written  as
\beq
\Pi  &=& \frac{1}{  2\sqrt{2}} \begin{pmatrix} 
    A & H \\ H^\dagger  & -A \\ 
      \end{pmatrix}  \ .
\eeq
Here $A$ and $H$ are  two by two matrices, with
\begin{align}
  A  =  \begin{pmatrix} 
    a & \quad \\ \quad  & a \\ \end{pmatrix}  & &
  H  =  \begin{pmatrix} 
    h^0+ih_3&ih_1+h_2 \\  
  i h_1- h_2 & h^0-ih_3   \\ \end{pmatrix} \ . \end{align}
   Here $a$ is the electroweak singlet and the matrix $H$ must satisfy 
   \be \sigma_2 H- H^* \sigma_2=0\ ,\ee and the fields $h_i$ are real.
   
Using the gauged $SU(2)_L$ to eliminate the unphysical NGB's $h_{1,2,3}$, vacuum misalignment with respect to the weak gauge interactions can be 
parameterized by an angle $\theta = \vev{h_0}/\sqrt2f$.
   When $a=0$, the sigma field is 
   \be \Sigma=
    \begin{pmatrix}0&c&0&i s\\ -c&0&-i s&0\\
   0&i s&0&c\\ -i s&0&-c&0 \end{pmatrix}, \ee
  where $c=\cos\theta$ and $s=\sin\theta$. 
  Note that this is invariant under the  custodial $SU(2)_c$ generated by $R^a+Q^a$.

\subsubsection{Gauge Interactions}
\label{sec: SUSPgauge}
The gauge  couplings explicitly break the $SU(4)$ global symmetry .  The unique  two derivative term for the non-linear sigma model is
 \begin{equation}
  \frac{f^2}{4} \tr\abs{D_\mu\Sigma}^2
\label{kineticterm}
\end{equation}
where the covariant derivative of $\Sigma$ is given by
\begin{equation}
  \label{eq:cd}
  D\Sigma = \partial \Sigma -  \left\{ i g W^a (Q^a
   \Sigma + \Sigma Q^{aT})  + i g' (Y \Sigma + \Sigma Y^T)\right\}
\end{equation}
and $g, g^\prime$ are the couplings of  $SU(2)_L$ and $U(1)_Y$, respectively.  
Misalignment of the vacuum will give tree level gauge boson masses
\begin{align}
m_w^2 = 	\frac{g^2}{2}\,f^2 s^2 && m_z^2 = \frac{g^2 + g'^2}{2}\, f^2 s^2\ .
\end{align}
The gauge interactions will lead to a quadratically divergent one loop correction to the potential for $\Sigma$.
This divergence may be absorbed into a counterterm for the following interactions
\begin{equation}
\label{gaugecw}
 -c g^2 f^4\sum_a \tr \left[(Q^a \Sigma)(Q^a \Sigma)^*\right] -
 c  {g'}^2 f^4 \tr \left[(Y \Sigma)(Y \Sigma)^*\right] .
\end{equation}
Here $c$ is a UV sensitive constant which parameterizes the leading gauge contribution to the effective potential for the PNGBs.
Dimensional analysis gives $c$ of order 1, while a technicolor-like UV completion of the model suggests $c$ is positive, that is, the gauge contribution to vacuum alignment prefers vanishing gauge boson masses\footnote{The analogous term in the  effective low energy description of QCD gives the (positive) mass squared splitting between the $\pi^+$ and $\pi^0$, see also ~\cite{Piai:2004yb}.}.

\subsubsection{Yukawa Interactions}

We construct the Yukawa sector in a manner conducive to building a UV completion, following reference~\cite{Katz:2003sn}.  The  interactions are
\begin{equation}
  \label{topYukawa}
  \CL_{t}=  - i \lambda_1 f \Psi \Sigma \overline{\Psi} + \lambda_2 f  \,{
  q_3} \overline{Q} + \lambda_3  f \, T \overline{t}_3  + {\rm h.c.}
  \end{equation}
  where we have defined 
\be
\begin{array}{cc}
\Psi \equiv \left( \begin{array}{ccc}
Q& B & T \end{array} \right), & \overline{\Psi} \equiv \left( \begin{array}{c}
\overline{Q} \\ \overline{T} \\  \overline{B}  \end{array} \right) 
 \end{array}\ ,
\ee
to be $SU(4)$ quartets and we use a notation where all fermions are left handed Weyl fields.  Here, $Q$ is a doublet under $SU(2)_L$.  $T$ and $B$ are $SU(2)_L$ singlets.  In addition to $\Psi, \overline{\Psi} $ we also have all the usual three generations of quark and lepton fields, which we take to be $SU(4)$ singlets.
The left handed top quark will be a mixture of the charge 2/3 fields in $\Psi$ and the quark doublet
\be
q_3 \equiv \left( \begin{array}{cc}
q_t & q_b  \end{array} \right)  
\ee
while the left handed anti-top will be a linear combination of charge -2/3 fields in $\overline\Psi$ and the
singlet 
\be\bar t _3.\ee
Other interactions\footnote{One modification, for example, requires adding an additional coupling, e.g. $\lambda_4 f \bar b B$,   to generate the bottom Yukawa coupling with the seesaw mechanism. The top-bottom quark mass difference can be explained through a small value for $\lambda_4$.  Such a setup may generate non-trivial consequences for the precision electroweak observables, $R_b$ and $A_{LR}$, and b-factory physics.} can be included to generate the light quark and lepton Yukawa couplings.  

A spurion analysis reveals the collective symmetry breaking mechanism of this sector. In the absence of the  gauge couplings and couplings $\lambda_{1,2,3}$, the global symmetry is actually $SU(4)^3$, with the fields $\Psi, \overline{\Psi},$ and $\Sigma$ transforming under different $SU(4)$'s.  The $\lambda_1$ coupling breaks the $SU(4)^3$ to a diagonal  $SU(4)$.  The coupling $\lambda_{2}$ partially breaks the $SU(4)$ under which $\overline\Psi$ transforms   and $\lambda_3$ partially breaks the $SU(4)$ under which $\Psi$ transforms. 
Only the collective application of all three couplings completely breaks the global symmetries and allows the top Yukawa to be nonzero.  The quark fields have the following quantum numbers 
\begin{center}
\begin{tabular}{ccccc}
  & $SU(3)_c$ 	& $SU(2)_L$ & $U(1)_Y$ & $U(1)_a$ \\ 
$q_3$		& 3 		& 2 		& 1/6 & -1 \\ 
 $\overline{t}_3$ 	& $\bar3$	& 1		& -2/3 & -1  \\ 
  \\
  $Q$& 3 		& 2 		& 1/6 & -1 \\
$B$ & 3		& 1		&-1/3 & 1  \\ 
 $T$ & 3 		& 1 		& 2/3 & 1
\end{tabular}
\end{center}
where $U(1)_a$ is the subgroup of $SU(4)$  under which the electroweak singlet transforms as a Goldstone boson.  The charge $2/3$ quark mass matrix
\begin{center}
\begin{tabular}{r|ccc}
    & $\overline{Q}_t$ & $\overline{T}$ &$\overline{t}_3 $ \\ \hline
$Q_t$& $ i\lambda_1 f \cos{\theta}$& $-\lambda_1 f \sin{ \theta}$&0 \\
$T$&$  - \lambda_1 f \sin{ \theta}$&
$ i\lambda_1 f  \cos{\theta}$& $\lambda_3 f$
\\ $q_3$ & $\lambda_2 f$ & 0 &0 \\ 
\end{tabular}
\end{center}
depend on the Higgs' vacuum expectation value 
\be
\theta = \langle h\rangle/ (\sqrt{2} f).
\ee
In the small $\theta$ limit, the effective theory contains the usual top Yukawa coupling to the Higgs doublet.
To find this Yukawa coupling, we expand equation~\ref{topYukawa} to first order in the Higgs
\beq
\CL_t &\supset&   -i f (-\lambda_1 Q_t + \lambda_2 q_3) \overline{Q}_t 
- i  f T(-\lambda_1 \overline{T} + \lambda_3 \overline{t}_3) 
\\ \nonumber 
&-& \frac{1}{\sqrt{2}}  \lambda_1 Q_t h \overline{T} 
- \frac {1}{ \sqrt{2}}  \lambda_1 T h \overline{Q}_t. 
\eeq
Here $\overline{Q}_t$ marries the linear combination $(-\lambda_1 Q_t + \lambda_2 q_t) /(\lambda_1^2 + \lambda_2^2)^{1/2}$. $T$ marries $(-\lambda_1 \overline{T} + \lambda_3 \overline{t}_3)/(\lambda_1^2 + \lambda_3^2)^{1/2}$
to become massive.  These heavy quarks are integrated out.  In the limit $v/f\ll1$, the light combinations
\begin{align}
t_L \equiv \frac{\lambda_2 Q_t + \lambda_1 q_t}{ \sqrt{\lambda_1^2 + \lambda_2^2}}  & & \overline{t}_R \equiv 
\frac{\lambda_3 \overline{T} + \lambda_1 \overline{t}_3}{\sqrt{\lambda_1^2 + \lambda_3^2}} 
 \end{align}
have the mass term 
\be\label{topmasterm}
\frac{\lambda_t}{\sqrt2}  \langle{ h \rangle} t_L\overline{t}_R \ee
and  the top Yukawa coupling is\footnote{Note that the factor of $\sqrt2$ in equation \ref{topmasterm} comes from the different conventions used to normalize real and complex scalar fields.}
\be
\lambda_t = \frac{ \lambda_1 \lambda_2 \lambda_3 }{ \sqrt{\lambda_1^2 + \lambda_2^2}\sqrt{\lambda_1^2 + \lambda_3^2} }\ .
\label{eq:topyuk}
\ee
In the intermediate limit, this treatment of the top Yukawa is not adequate, but the requirement of all three couplings to generate a mass for the top quark holds for any value of $\theta$, as can be seen by noting that if any one of $\lambda_{i}$ vanishes so does the determinant of  the charge 2/3 quark mass matrix.

\subsection{A Potential for the Singlet}

The Yukawa and gauge interactions, equations~\ref{eq:cd} and~\ref{topYukawa}, preserve a global  $U(1)_a$, under which the electroweak singlet  $a$ transforms as a Nambu-Goldstone boson. If massless, this singlet would be ruled out by searches for $K\rightarrow \pi\, + a$. This symmetry is broken by the QCD anomaly, but this will not give $a$ a sufficiently large mass.  To increase the mass of $a$ we add a symmetry breaking term 
\be
\label{massspurion}
\CL_m = - f^3 \tr[ M^\dagger \Sigma ] + {\rm h.c.}
\ee
where 
\be
M\equiv m \Sigma_0 \ .
\ee
This term provides a potential for both $a$ and $h$.
We assume this term is sufficiently large to avoid experimental constraints on $a$ and sufficiently small so that electroweak symmetry breaking can occur without fine tuning.

We will take $m$ to be real and positive\footnote{Note that a more general choice of the matrix $M$ will not affect our conclusions, as long as the spurion is gauge invariant.  Phases in $M$ can simply be absorbed by redefining $\Sigma$. Allowing the eigenvalues to be unequal is possible, but does not significantly change the low energy phenomenology, so we use equal eigenvalues for simplicity.}.  If $m$ is large enough, the Higgs will dominantly decay into $a$ pairs.  This non-trivial decay rate has profound consequences for collider signatures, and could affect the bounds from the Large Electron Positron Collider (LEP) on the Higgs mass.  To this order, the scalar potential is nearly invariant under $a\rightarrow -a$. The interactions \ref{topYukawa} give the $a$ boson an axion-like coupling to quarks and gluons, and  $a$ will decay to two quarks or two gluons.  Whether of not a significant fraction of these decays will be into $b$ quarks is a model dependent question \cite{hep-ph/0411213}. Depending on the $a$ mass, the $a$ particle will predominantly manifest as either one jet or two jets. We discuss some consequences in section \ref{sec: coll}.

\subsection{Vacuum Alignment, the Higgs Mass, and Fine Tuning}
\label{sec:effpot} 
  
In the subsections above, we detailed the  interactions interactions needed to break $SU(4)$ global symmetry and select a preferred alignment for the new vacuum.  In this section, we consider the Higgs effective potential at one loop.    The radiative corrections from the top  sector dominate, being  enhanced by color, fermion and symmetry multiplicities as well as an order one top Yukawa coupling.

In general, the  one loop radiative corrections to the potential may be divided into quadratically divergent, logarithmically divergent, and soft contributions. 
The divergent contributions from the fermion sector  are respectively
\be
-\frac{3 \Lambda^2}{ 8 \pi^2} \tr M^\dagger\!M,
\label{yukquaddiv}
\ee
\be
-\frac{3  }{16 \pi^2} \tr(M^\dagger\!M)^2\ln(M^\dagger\!M/\Lambda^2),
\label{yuklogdiv}
\ee
 where $M$ is the fermion mass matrix. 
Note that since the Yukawa interaction does not break $U(1)_a$, the Yukawa contribution to the effective potential is independent of $a$.  
Note also that
\begin{align}
 \frac{\partial}{\partial \theta}{\rm Tr} M^\dagger M=0 & &
\frac{\partial}{\partial \theta}{\rm Tr} (M^\dagger M)^2=0
 \label{nodiv}
 \end{align}
This ensures that except for a constant term,  the one loop contribution to the potential from this sector is  insensitive  to the cutoff,  and is given by the expression
\be\label{topHiggs}
-\sum_i \frac{3 |m_i^2|^2}{ 16 \pi^2} \log |m_i^2|
\ee
where $|m^2_i|$ are the eigenvalues of the matrix  $m_{2/3} m_{2/3}^\dagger$ .

Besides the top sector, there are other radiatively generated corrections to the Higgs's potential.  These contributions are parametrically not as large as top contributions and are UV sensitive.  We can analyze the vacuum alignment by minimizing all of the contributions to the potential
\be
V \sim  f^4 \left(G(g^2, c,\theta) + F( \lambda_1, \lambda_2, \lambda_3, \theta) + E(m/f, \theta, a/f)\right)
\ee
where $G$, $F$ and $E$ are respectively the leading contributions from the gauge interactions, Yukawa interactions, and the spurion of eq. \ref{massspurion}.  The most important gauge contributions to the vacuum alignment are eq. \ref{gaugecw}.  The UV sensitivity of the one-loop correction is absorbed by the unknown parameter $c$,  which is expected to be positive and of order 1.  Similarily,  the spurion sector contains a small explicit symmetry breaking parameter $m$ which will play a role in determining the Higgs' mass and in determining the nature of the fine-tuning in this theory.  Note that the contributions $G$ and $F$ are independent of $a$, and $E$, and therefore $V$, is clearly minimized  for $a=0$. We thus set $\vev{a}=0$ for the remainder of the analysis. 

This model contains a physical Higgs boson, whose couplings to the gauge bosons approach those of the Standard Model Higgs in the $\theta\rightarrow 0$ limit. The traditional Higgs search, e.g. at LEP,  which assumes Higgs production via the $\mathrm{Higgs}\!-\!\mathrm{Z}\!-\!\mathrm{Z}$ coupling  should therefore place constraints on this model. The mass of the physical Higgs boson depends on $f$, the $\lambda_i$ parameters, and $m$. The top mass and the heavy quark masses are a  function of $f$, and the $\lambda_i$ parameters. 
We find that the Higgs mass in this model is typically below 120 GeV.
 To compute the  Higgs  mass to leading order, we choose $f$, and fix $\theta$ so that the $W$ and $Z$ masses come out to their observed values. We then fix the ratios $\lambda_2/\lambda_1$ and $\lambda_3/\lambda_1$ to some typical  values of order one, numerically find the eigenvalues of the
charge 2/3 mass matrix, and determine $\lambda_1$ so that the lightest eigenvalue is at 180 GeV.  
We then choose some values for $m$ and compute  $c$ so that the potential is indeed minimized at the necessary value of $\theta$. We then compute the Higgs mass by numerically computing the second derivative of the Higgs potential with respect to $\theta$ at the minimum.
 \FIGURE[t]{ 
 \centerline{\epsfxsize=4.5 in \epsfbox{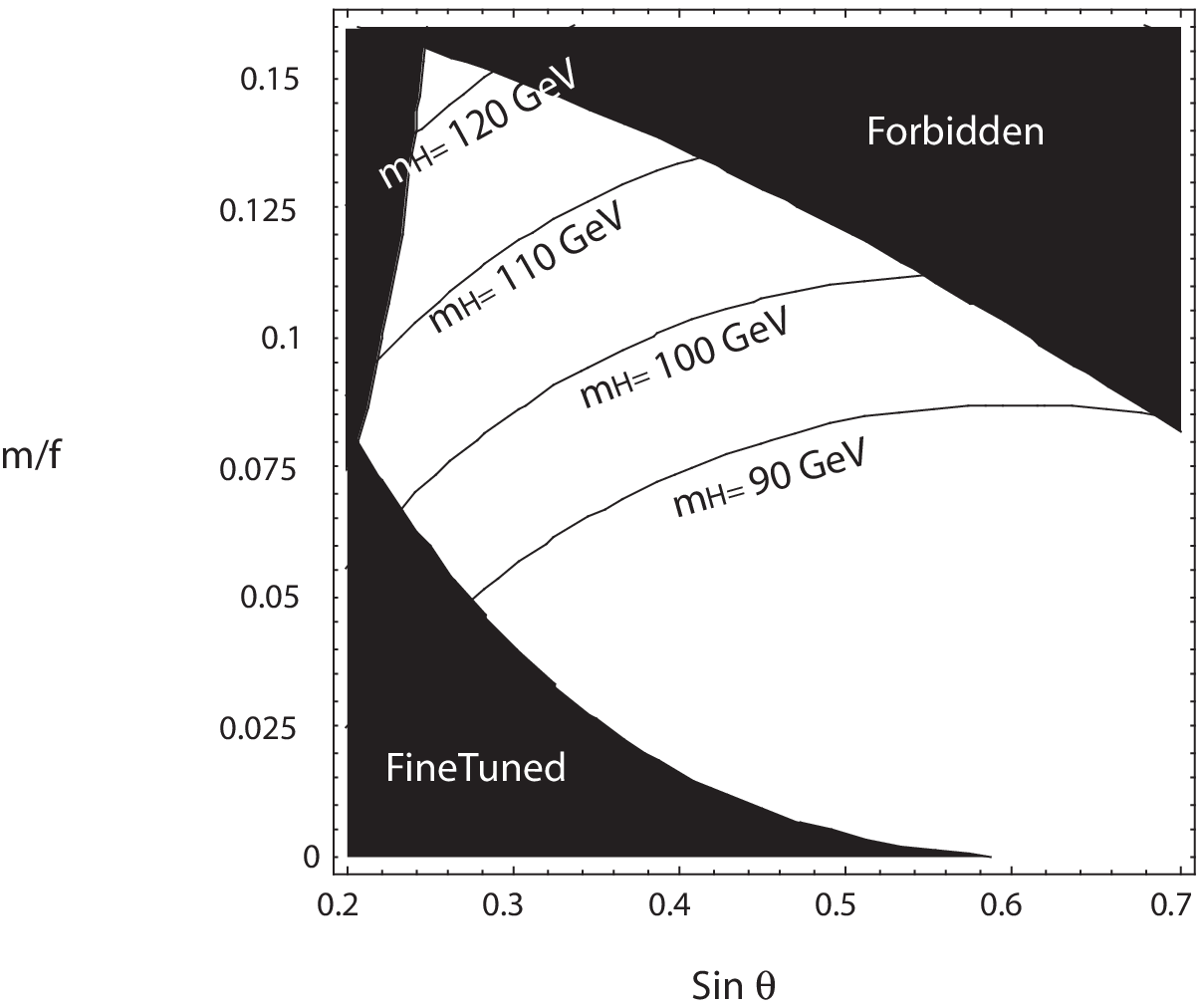}  } 
 \noindent\caption{Contour Plot of the Higgs mass in SU(4)/SP(4) model,  with the top mass fixed at 180 GeV, and $\lambda_1/\lambda_3$ and $\lambda_1/\lambda_2$ fixed to 1, for various values of $\sin\theta\equiv v/\sqrt2f$ and the symmetry breaking spurion $m$. The black regions either have negative values for the parameter $c$ or require more than 10\% finetuning of the parameters $c$ or $m$.}}
To check for finetuning, we compute the sensitivity of $\sin\theta$ to the values of the free parameters $c$ and $m$, holding all other parameters fixed.  If small changes in $c$ or $m$, lead to large changes in the minimum of the potential we then have to finely tune these parameters to achieve the observed weak boson masses.   We thus evaluate  numerically ``sensitivity parameters'' $S_c$ and $S_m$ \cite{Barbieri:1987fn}
 \be
 S_c\equiv \frac{\partial \log \sin\vev{\theta}}{\partial \log c}
 \ee and
  \be S_m\equiv \frac{\partial \log \sin\vev{\theta}}{\partial \log m}\ .
\ee
  The amount by which the parameters $c$ and $m$ have to be finely tuned in order to produce the observed $W$ mass is  of order $1/S_{c,m}$. Values of $S_{c,m}$ which are greater than 10 will be regarded as indicative of unacceptable fine tuning. In principle we should also check for sensitivity with respect to all the parameters, however when the minimum of potential is not excessively sensitive to $m$, $c$ we do not  expect it to be excessively sensitive to  other parameters such as the $\lambda_i$ either.  
Note that $c$ and $m$ are renormalized parameters. 

Since a primary goal of the IH is to eliminate sensitivity to short distance physics, and since the weak scale has order one sensitivity to the parameter $c$,  we also should check the sensitivity of the renormalized parameter $c$ to short distance physics. To check the sensitivity of $c$ to short distance physics would require knowledge of the UV physics\footnote{ As a crude check, one could simply use a naive cutoff as a toy model of short distance physics, 
as was done in ref. \cite{casas}, and check for sensitivity to the value of the cutoff. }. One model would be to eliminate such sensitivity by including additional gauge bosons, as in \cite{hep-ph/0105239}, and check for sensitivity to the additional gauge boson masses. The results  suggest that such sensitivity is less than 10 provided the new gauge boson masses are less than 6 TeV. This suggests that avoiding finetuning implies additional physics at a scale below 6 TeV.  In any case, the nonrenormalizable nature of our effective theory implies new physics below a scale of approximately $4\pi f$.  We list some ideas for such a UV completion in section \ref{sec: UV}. Similar statements apply to the other parameters
on which the weak scale depends, such as $\lambda_i$.
 
\FIGURE[t]{ 
 \centerline{\epsfxsize=4.5in \epsfbox{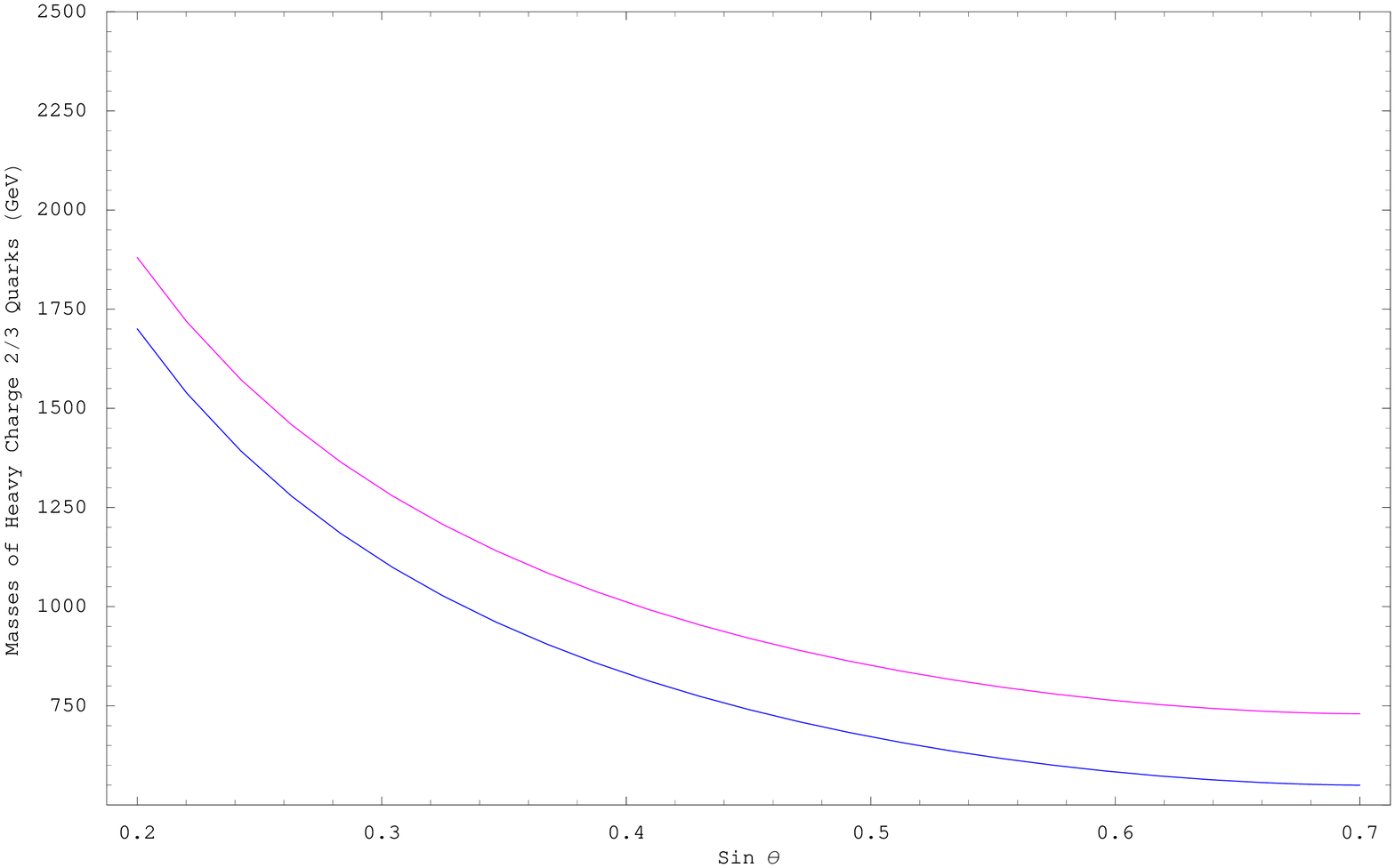}  } 
 \noindent\caption{Masses (GeV) of heavy charge 2/3 quarks in the SU(4)/SP(4) model, as a function of $\sin\theta\equiv v/\sqrt2f$, with $\lambda_1/\lambda_3$ and $\lambda_1/\lambda_2$ fixed to 1.}  }
 
We find that when $m$ is small, for any values of $\sin\theta$ other than 0 or 1, the sensitivity parameter $S_c$ is greater than 10, while $S_m$ is negligibly small. Since the parameter $c$ is then calculable in terms of $\theta$, the Higgs mass is predicted in this limit in terms of $\theta$ and the ratios $\lambda_2/\lambda_1$ and $\lambda_3/\lambda_1$ . We find that  in this limit the Higgs mass is mostly sensitive only to the mass of the next lightest charge 2/3 quark, the $T'$, and is lighter than 114 GeV unless the next lightest quark is heavier than about 9 TeV.  When the Higgs is heavier than 114 GeV,  $S_c$ is always above $260$. The reason for the light Higgs and 0.4\% finetuning is that the $\theta$ dependence of the functions $G$ and $F$ is very similar, unless the logarithms in $F$ are large. For small $\theta$ both $G$ and the largest terms in $F$ are proportional to $\sin^2\theta$. We conclude that if we are to avoid finetuning, we  must assume that $m$ is large enough to  have an effect on the alignment which is comparable to the effects of the gauge and Yukawa couplings.
When $m$ is moderately large,  the function $E$, which is proportional to $\cos\theta$,  plays a role in determining the alignment. For a range of values of $m$,  intermediate $\theta$ is easy to obtain without fine tuning. For non-finely tuned values of the parameters, the Higgs mass is between 90 and 130 GeV. 

 In figure 1 we  give a contour plot of the 
 Higgs mass as a function of $m$ and $\sin\theta$, in the region where the  the sensitivity parameters $S_{c,m}$ are less than 10 and $c$ is positive,   with the top mass fixed at 180 GeV, and $\lambda_1/\lambda_3$ and $\lambda_1/\lambda_2$ fixed to 1. For the same ratios of $\lambda_1/\lambda_3$ and $\lambda_1/\lambda_2$, in figure 2 we show the masses of the new charge 2/3 quarks, as a function of $\sin\theta$.
We see that   when the $\lambda$'s are equal, intermediate values of $\theta$   allow for a Higgs mass in the 100-120 GeV range without finetuning. We have checked that these results are robust against order one variations in the ratios of 
$\lambda_{1,2,3}$, with the  Higgs mass always below 130 GeV in non-finetuned regions of parameter space.
\smallskip

\subsection{The $SU(5)/SO(5)$ Model}
This model has 14 PNGBs, transforming under $SU(2)_L \otimes SU(2)_R$ as
\be (2,2)+(3,3)+ (1,1)\ee and
$SU(2)_L \otimes U(1)_Y$ as
\be
 1_0 \oplus 2_{\pm1/2}\oplus 3_0\oplus 3_{\pm1}\ .
\ee
The presence of  doubly charged scalar bosons at the weak scale is an interesting feature of this model, which could be an important clue into the composite nature of the Higgs.
The PNGBs may be described by a 
symmetric unitary $5\times 5$ matrix valued field  $\Sigma$, which transforms as $\Sigma \to V \Sigma V^T$
under $SU(5)$ transformations $V$. 
\subsubsection{Gauge Interactions}
 The $SU(2)_L$ electroweak  generators are embedded into $SU(5)$ as
\begin{align}
  Q^a &=
  \begin{pmatrix}
    \sigma^a/2 &\quad&\quad\\ \quad&\quad&-{\sigma^a}^*/2
  \end{pmatrix}, \\
  \intertext{while the generators of the $ U(1)_Y$ are given by}  Y &=
    \diag(1,1,0,-1,-1)/2 \ .\end{align}
    The generators of the approximate $SU(2)_R$ are
    \begin{align}
     R^1=\begin{pmatrix} \quad  & \quad &-i \sigma_2/2 \\ \quad & 0 &\quad \\  i\sigma_2/2
    &\quad&\quad\end{pmatrix}  &&
 R^2=\begin{pmatrix} \quad  & \quad & -\sigma_2/2 \\ \quad & 0 &\quad \\ -\sigma_2/2
    &\quad&\quad\end{pmatrix}  &&   R^3=Y
 \end{align}
 \subsection{Effective Sigma Model}
The $SU(2)_L \otimes U(1)_Y$ preserving vacuum $\Sigma_0$  is
\begin{equation}
  \label{sigma0}
  \Sigma_0=
  \begin{pmatrix}  \qquad  & \qquad & \openone \\ \qquad & 1 &\qquad \\ \openone
    & \qquad& \qquad
  \end{pmatrix}
  \ .
\end{equation}
The Nambu-Goldstone bosons are fluctuations about this background in the direction of the broken generators, 
$\Pi \equiv \pi^a X^a$.  They are parameterized as 
\be
  \Sigma(x) = e^{i \Pi/f} \Sigma_0 e^{i \Pi^T/f} =  e^{2 i \Pi/f}
  \Sigma_0 .
\ee
In this basis, the PNGB matrix may be written
\begin{equation}
  \label{pgb} 
  \Pi=
  \begin{pmatrix}
    \frac{\eta}{\sqrt{40}}\one+\frac{\vec\phi\cdot \vec\sigma}{\sqrt8}
    &\frac{h^T}{2}&\frac{\tilde\phi}{\sqrt2}\\
    \frac{h^*}{{2}}&-\frac{2\eta}{\sqrt{10}}
    &\frac{\tilde{h}}{{2}}\\ \frac{{\tilde\phi}^\dagger}{\sqrt2}&\frac{\tilde{h}^\dagger}{{2}}&
   \frac{\eta}{\sqrt{40}}\one+\frac{\vec\phi\cdot {\vec\sigma}^*}{\sqrt8}
 \end{pmatrix}
 \end{equation}
  where $\tilde\phi$ is a complex symmetric two by two matrix describing an electroweak triplet with hypercharge one,
  \be\tilde\phi=\begin{pmatrix}\tilde\phi^{++}&\frac{\tilde\phi_+}{\sqrt2}\\
  \frac{\tilde\phi_+}{\sqrt2}&\tilde\phi_0\end{pmatrix}\ee and $\vec\phi$ describes a real electroweak triplet with hypercharge 0. Note that the 6 components of $\tilde\phi$ and 3 components of $\vec\phi$  together transform as a (3,3) under $SU(2)_R\otimes SU(2)_L$.  The remnant custodial $SU(2)_c$ generated by $R^a+Q^a$ will be preserved in a vacuum where the only PNGB's with vevs are $h^0$, $\eta$, and $\vev{\phi_3}=\vev{\tilde\phi_0}$. Note  that small spurions which do not arise from any gauge interactions will be required to provide a potential for the   $\eta$, which does not get a mass from the gauge interactions\footnote{In the ``Littlest'' Higgs model \cite{Arkani-Hamed:2002qy} this $SU(2)_L \otimes U(1)_Y$ was gauged and the associated NGBs were eaten. However such a  new gauge symmetry provides an explicit source of $SU(2)_R$ breaking and can lead to difficulties with precision electroweak corrections.}.  Assuming preservation of the $SU(2)_c$ generated by $R^a+Q^a$,  the alignment of the vacuum  $SU(2)_L \otimes U(1)_Y$ preserving direction may be parameterized by three angles $\theta, \hat \eta,$ and $\hat \phi$, and sigma field
\begin{equation}
  \label{sigma}
  \Sigma=
  \begin{pmatrix}  0\quad & 0& 0& e^{2i(\hat\phi+ \hat\eta)}  &\quad0 \\ 
  0\quad&\frac12 e^{-6 i\hat\phi+ 2 i \hat\eta} (-e^{8 i \hat\phi} + \cos2\theta)& \frac{i}{\sqrt2}e^{-3 i(\hat\phi+\hat\eta)} \sin2\theta& 0& \frac12 e^{-6 i\hat\phi+ 2 i\hat \eta} (e^{8 i \hat\phi} + \cos2\theta)\\
0\quad&\frac{i}{\sqrt2} e^{-3 i(\hat\phi+\hat\eta)} \sin2\theta& e^{-8 i\hat\eta} \cos (2\theta)& 0 & \frac{i}{\sqrt2} e^{-3 i(\hat\phi+\hat\eta)} \sin2\theta\\
   e^{2 i(\hat\phi+\hat\eta)} &0&0&0&\quad0\\
    0\quad& \frac12 e^{-6 i\hat\phi+ 2 i \hat\eta} (e^{8 i \hat\phi} + \cos2\theta)&\frac{i}{\sqrt2} e^{-3 i(\hat\phi+\hat\eta)} \sin2\theta &0& \frac12 e^{-6 i\hat\phi+ 2 i\hat\eta} (-e^{8 i\hat\phi} + \cos2\theta) \end{pmatrix}
  \ .
\end{equation}

 As in the $SU(4)/SP(4)$ model, in the small angle limit, $\theta=\vev{h}/(\sqrt2f)$.
To leading order in the weak couplings, the masses of the $W$ and $Z$ bosons are
\be
m^2_W=\frac{g^2}{4}f^2(1-\cos8\hat\phi \cos2\theta),\quad\quad
m^2_Z=\frac{g^2+{g'}^2}{4}f^2(1-\cos8\hat\phi \cos2\theta)\ .
\ee
Note that in this model $SU(2)_L \otimes U(1)_Y$ could be broken by  triplet vevs  while retaining the tree level relationship $m_W=\cos\theta_w m_Z$, due to the fact that proper alignment of the triplet vevs can still preserve a remnant custodial $SU(2)_c$. However, lowest energy vacuum alignment will have $\vev{\phi_3}=\vev{\tilde\phi_0}=0$.
 
 The gauge interactions break SU(5) explicitly. One loop renormalization 
of this  model requires the  introduction of spurions
\be\label{spurone} -c g^2f^4 \Tr Q\Sigma Q  \Sigma^\dagger + h.c.\ee
and \be \label{spurtwo}-c g'^2 f^4 \Tr Y\Sigma Y \Sigma^\dagger+ h.c. \ee
These spurions induce terms in the potential involving $\theta$ and $\tilde\phi$ 
which are minimized  in an $SU(2)_L\otimes U(1)_Y$ preserving direction.
\subsubsection{Additional Potential Term}
In order to provide mass for the electroweak singlet, we introduce a custodial $SU(2)_c$ preserving spurion
\be  \label{massspur} \xi_1\Tr(\Sigma M) + h.c.
\ee with
\be M=\begin{pmatrix} \quad  & \quad & m_2\openone  \\ \quad & m_1 &\quad \\ m_2\openone 
    &\quad&\quad
  \end{pmatrix}
  \ .
\ee
In a UV complete model where  $\Sigma$ is a low energy effective description of a  fermion condensate,
the term $\ref{massspur}$ could arise from fermion mass terms.

Note that \ref{spurone}, \ref{spurtwo} and \ref{massspur} all give contributions to the potential whose $\theta$ dependence is proportional to $\cos2\theta=2 \sin^2\theta-1$. Such a potential can only stabilize $\theta$ at 0 or $\pi/2$, mod $\pi$. Motivated by the assumption of a technicolor-like UV completion, we assume that signs of the parameters are such that these contributions to the potential are minimized at $\theta=\tilde\phi=\vec\phi=0$.

\subsubsection{Yukawa Interactions}
As In the $SU(4)/SP(4)$ model,  we will   rely on UV insensitive radiative corrections from the top sector to align the vacuum in an $SU(2)_L\otimes U(1)_Y$ breaking direction. No finetuning will be required to obtain an intermediate value of $\theta$, with approximate custodial $SU(2)_c$ preservation and   a sufficiently heavy Higgs.

In  ref. \cite{Arkani-Hamed:2002qy}, only a single additional quark is introduced in order to generate the top quark mass.  However this mechanism produced hard breaking of the custodial $SU(2)_c$, and required an additional custodial $SU(2)_c$ breaking spurion  in  the Higgs potential.  The presence of the light triplets makes breaking of custodial $SU(2)_c$ quite dangerous, as the resulting potential could produce custodial $SU(2)_c$ violating triplet vevs, and change the $\rho$ parameter at tree level.
We therefore follow ref. \cite{Katz:2003sn}, with additional quarks in complete SU(5) multiplets,  in order to generate the top quark Yukawa coupling. 
That is, we add quarks $\chi\sim(Q,T, P)$ and $\bar\chi\sim(\bar P, \bar T, \bar Q)$ in complete SU(5) multiplets, which get mass from the $\Sigma$ field. In a fermion condensate model of the $\Sigma$, such quarks might be  composites of the new fermions.  The new quarks have the following gauge quantum numbers.
\begin{center}
\begin{tabular}{cccc}
  & $SU(3)_c$ 	& $SU(2)_L$ & $U(1)_Y$  \\ 
$Q$		& 3 		& 2 		& 1/6 \\ 
 $T$ 	& 3	& 1		& 2/3  \\ 
$P$ & 3		& 2		&7/6 
\end{tabular}
\end{center}
The top mass arises from the collective effects of three terms:
\be\lambda_1 f \bar\chi\Sigma\chi+ \lambda_2 f \bar{t}_3 T + \lambda_3 f \bar{Q} q_3+ h.c.\ee
The charge $2/3$ quark mass matrix is
\be
m_{2/3}=\lambda_1 f\begin{pmatrix} 
\frac12 e^{-6 i\hat\phi+ 2 i \hat\eta} (-e^{8 i \hat\phi} + \cos2\theta)& \frac{i}{\sqrt2}e^{-3 i(\hat\phi+\hat\eta)} \sin2\theta&  \frac12 e^{-6 i\hat\phi+ 2 i\hat \eta} (e^{8 i \hat\phi} + \cos2\theta)&0\\
\frac{i}{\sqrt2} e^{-3 i(\hat\phi+\hat\eta)} \sin2\theta& e^{-8 i\hat\eta} \cos (2\theta)&  \frac{i}{\sqrt2} e^{-3 i(\hat\phi+\hat\eta)} \sin2\theta&0\\
  \frac12 e^{-6 i\hat\phi+ 2 i \hat\eta} (e^{8 i \hat\phi} + \cos2\theta)&\frac{i}{\sqrt2} e^{-3 i(\hat\phi+\hat\eta)} \sin2\theta &\frac12 e^{-6 i\hat\phi+ 2 i \hat\eta} (-e^{8 i \hat\phi} + \cos2\theta) &\lambda_3/\lambda_1\\
 0&\lambda_2/\lambda_1&0&0\end{pmatrix}\ee
 One eigenvalue of this matrix always has absolute value $\lambda_1 f$. In the small $\theta$ limit, the other masses are $\sqrt{\lambda_1^2+\lambda_2^2} f$, $\sqrt{\lambda_1^2+\lambda_3^2} f$, and
 $\lambda_1\lambda_2\lambda_3/(\sqrt{\lambda_1^2+\lambda_2^2}\sqrt{\lambda_1^2+\lambda_3^2} )(\theta f)$.
   
    \FIGURE[t]{ 
 \centerline{\epsfxsize=6.5 in \epsfbox{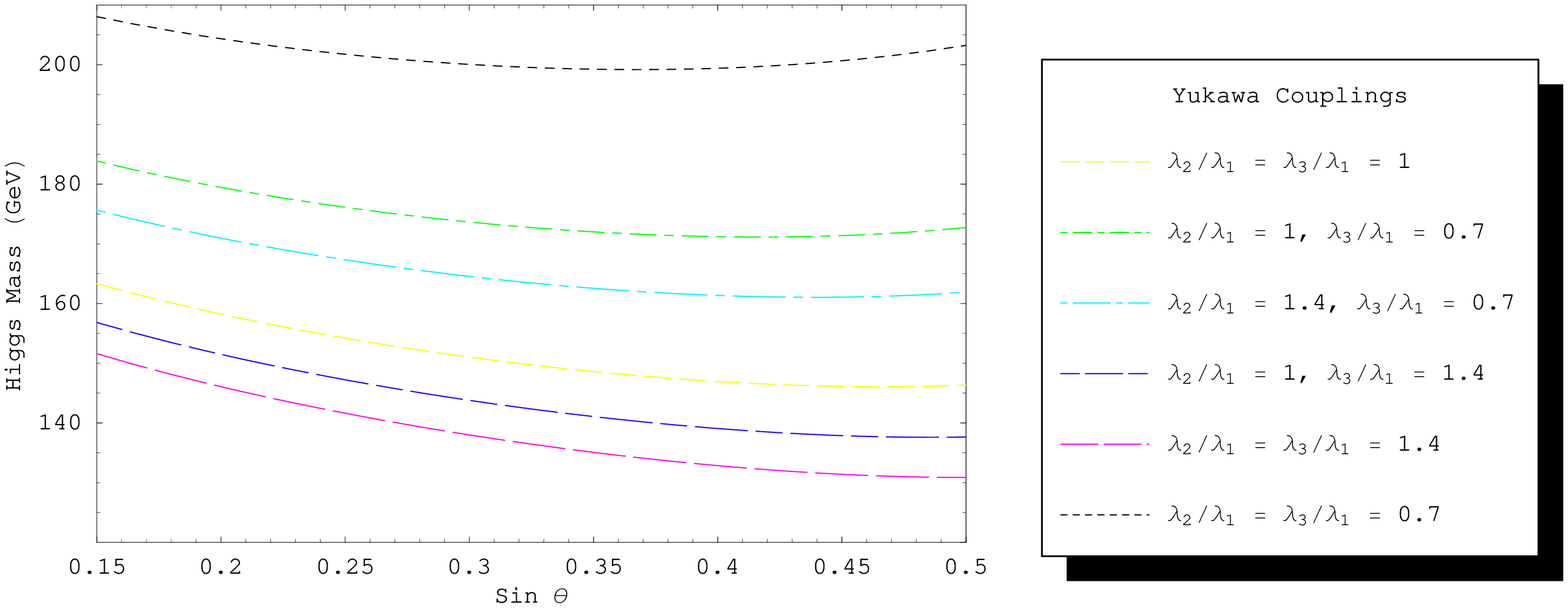}  } 
 \noindent\caption{ 
 Plot of the Higgs mass in SU(5)/SO(5) model,  with the top mass fixed at 180 GeV, and various values of $\lambda_3/\lambda_1$ and $\lambda_2/\lambda_1$, as a function of  $\sin\theta\equiv v/\sqrt2f$. 
 }
 }
 
\subsubsection{Vacuum Alignment}
As in the $SU(4)/Sp(4)$ model, 
$\Tr m_{2/3} m_{2/3}^\dagger$  and $\Tr( m_{2/3} m_{2/3}^\dagger)^2$ are both independent of the vacuum alignment and so
the one loop correction to the scalar potential from this sector is UV insensitive. 
Furthermore, note that the eigenvalues of the matrix $ m_{2/3} m_{2/3}^\dagger$ depend only on $\theta$. The corrections to the effective potential from the top sector favor  a nonvanishing $\theta$, but are insensitive to  $\hat \phi$ and  $\hat\eta$.  One could see this by rescaling $\chi$ by $\Sigma^\dagger$, followed by phase rotations on $\bar{t}_3$,$\bar T$, and $T$, for instance. Since the gauge interactions and spurion terms favor zero for all the PNGB vevs,  only $\theta$ is  nonvanishing  in the ground state. 
To find the mass of the physical  Higgs boson, we use the same procedure as in section 
\ref{sec:effpot}. We do not obtain the same results, because, beyond quadratic order in $\theta$, the dependence of the gauge boson masses and charge 2/3 quark masses on $\theta$ is not the same in the $SU(5)/SO(5)$ model as they are in the $SU(4)/SP(4)$ model. Also, because the gauge and mass spurion terms are proportional to the same function of  $\theta$, there is only one free parameter in the determination of the alignment, other than the $\lambda_{1,2,3}$ parameters.  A linear combination of $c$ and $m$, which we call $c'$,  is fixed by the value of $\theta$. Thus the Higgs mass is calculable as a function of $f=v/(\sqrt2\sin\theta)$ and the heavy quark masses.  We find that  no finetuning is required to obtain a minimum of the potential at an intermediate value of $\theta$ with a sufficiently heavy Higgs. The sensitivity parameter $S$ is defined to be 
\be S\equiv \frac{\partial \log m_w}{\partial \log c'}  \ . \ee

\FIGURE[t]{ 
 \centerline{\epsfxsize=6.5 in \epsfbox{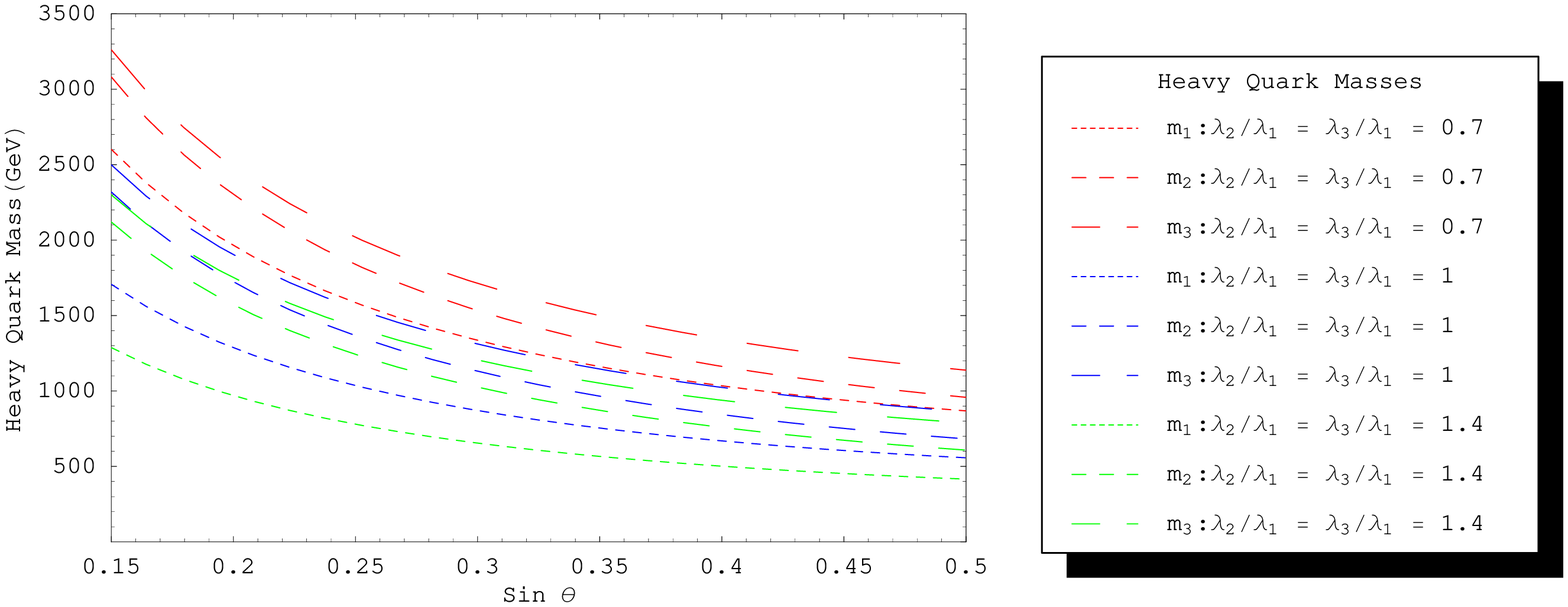}  } 
 \noindent\caption{Masses in GeV of the heavy charge 2/3 quarks in the $SU(5)/SO(5)$ model, as a function of $\sin\theta\equiv v/\sqrt2f$, for several values of $\lambda_3/\lambda_1$,$\lambda_2/\lambda_1$.}  
 }

In Figures 3-5  we plot the Higgs mass, the heavy charge 2/3 quark masses, and  $S$, as a function of $\sin\theta$, for several different values of the ratios $\lambda_{2,3}/\lambda_1$, with the top mass fixed to 180 GeV. The masses of the $\vec\phi$, $n$, and $\tilde\phi$ scalars are not predicted, as these depend on the eq.~\ref{massspur} parameters, and not just on the combination $c'$. Note that the leading terms in the resulting potential are even under the $Z_2$ transformations $\vec\phi\rightarrow -\vec\phi$, $\eta\rightarrow-\eta$, and $\tilde\phi\rightarrow-\tilde\phi$.  These symmetries are broken, however, by the couplings to the quarks, and so all the new scalars are unstable, with the lightest decaying dominantly into hadrons.

\FIGURE[t]{ 
 \centerline{\epsfxsize=6.5 in \epsfbox{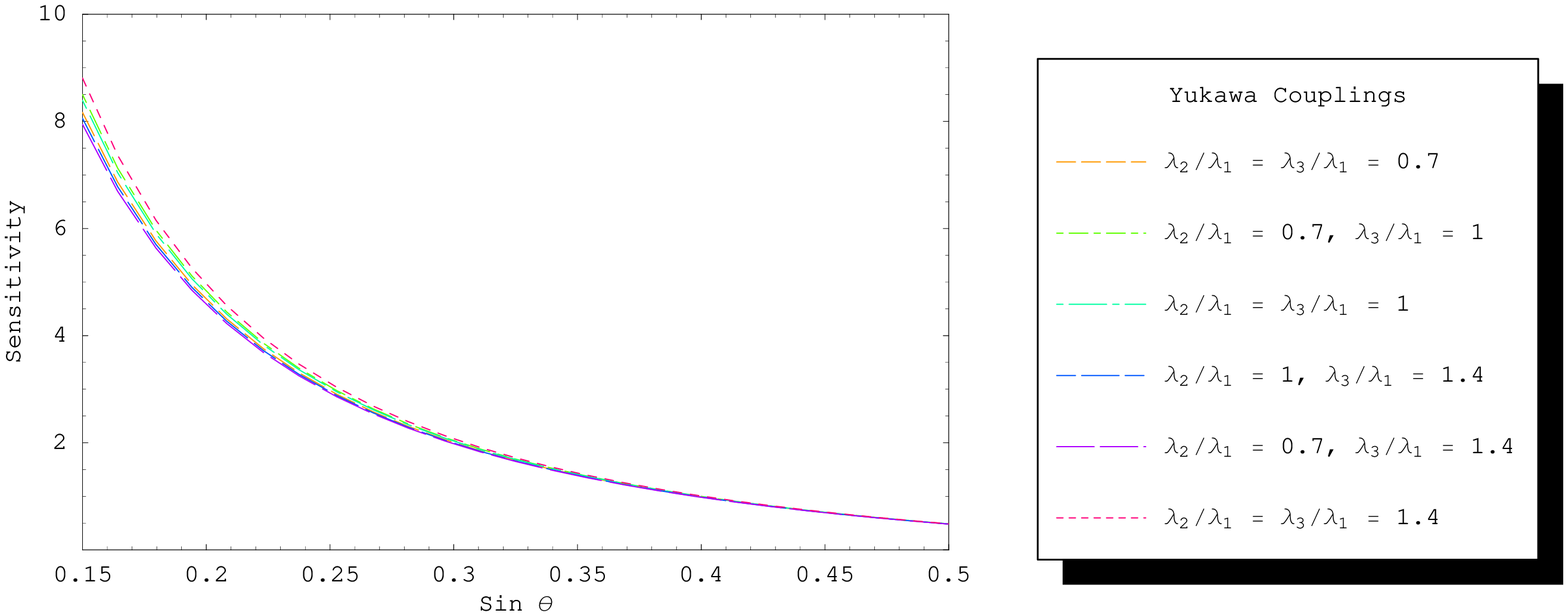}  } 
 \noindent\caption{Sensitivity of electroweak symmetry breaking to the $c'$ parameter in the $SU(5)/SO(5)$ model, as a function of $\sin\theta\equiv v/\sqrt2f$, for $(\lambda_2/\lambda_1, \lambda_3/\lambda_1)=(0.7,0.7),(0.7,1),(1,1),(1,1.4),(0.7,1.4),(1.4,1.4)$. Note that the amount of sensitivity mainly depends on the value of $\sin\theta$, and that for $f\lsim 1.2$~TeV the required finetuning of the $c'$ parameter is no worse than 12\%.}
 }  

\subsection{The $SU(6)/Sp(6)$ Model}

This model allows for two Higgs doublets, while preserving the Standard Model prediction for the $\rho$ parameter at tree level. The nonlinearly realized global symmetry is $SU(6)$, with an $SP(6)$ preserved by the vacuum. In analyzing the preservation of $SU(2)_c$ by the vacuum alignment it is convenient to focus on an $SU(2)_L \otimes SU(2)_{R1} \otimes SU(2)_{R2}$ subgroup of the $SU(6)$, where $SU(2)_L$ is the electroweak gauge group, and the custodial symmetry $SU(2)_c$ is the  diagonal subgroup of $SU(2)_L \otimes SU(2)_{R1} \otimes SU(2)_{R2}$.
In a reference vacuum with $SU(2)_L \otimes SU(2)_{R1} \otimes SU(2)_{R2}$ preserved,  this model has 14 PNGBs transforming under $SU(2)_L \otimes SU(2)_{R1} \otimes SU(2)_{R2}$ as
 \be
 (2, 2, 1) \oplus (2, 1, 2) \oplus (1, 2, 2) + (1, 1, 1) + (1, 1, 1)
 \ee 
 To recover the Standard Model hypercharge assignments, we gauge the $T_3$ generator within each custodial $SU(2)_c$ yielding the following PNGB charge assignments under $SU(2)_L \otimes U(1)_Y$:
 \be
 2_{\pm 1/2} \oplus  2_{\pm 1/2} \oplus  1_1 \oplus 1_{-1} \oplus 1_0 \oplus 1_0 \oplus 1_0 \oplus 1_0
\ .  \ee
 These goldstone bosons are described by a sigma field transforming as $\Sigma \to V \Sigma V^T$ under $SU(6)$ transformations, $V$.  The reference sigma field $\Sigma_0$  is
  \be
 \Sigma_0 = \begin{pmatrix}
 			\sigma_2 & \quad & \quad \\
			\quad  & \sigma_2 & \quad \\
			\quad  & \quad  & \sigma_2
			\end{pmatrix} .
 \ee
 
 It is well known that models with two light Higgs doublets typically have large  flavor changing neutral currents from Higgs exchange, unless restrictions are placed on the couplings to the quarks and leptons. One simple way to avoid excessive FCNC is to have only one doublet couple to  fermions of each electric charge.  This can be guaranteed by a discrete or a continuous symmetry, which can either be unbroken or broken softly. 
 \subsubsection{Gauge Interactions}
 We embed the electroweak group as
 \begin{align}
Q^a = 	\half \begin{pmatrix}
				\sigma_a & \quad 	& \quad \\
				\quad	& \quad	& \quad \\
				\quad	& \quad	& \quad
				\end{pmatrix} && 
				Y = \mathrm{diag}(0, 0, 1, -1, 1, -1)/2
\end{align}
The two  $SU(2)_R$ generators are
\begin{align}
R_1^a = \half \begin{pmatrix}
											\quad	& \quad	& \quad \\
											\quad	& \sigma^a & \quad \\
											\quad 	& \quad	& \quad
											\end{pmatrix} 
&&  R_2^a = \half \begin{pmatrix}
											\quad	& \quad	& \quad \\
											\quad	& \quad & \quad \\
											\quad 	& \quad	& \sigma^a
											\end{pmatrix} 
\end{align}

The relations between the broken and unbroken generators are the same as in the $SU(4)/Sp(4)$ model, equations \ref{eq: SUSPunbroken} and \ref{eq: SUSPbroken}.   The PNGB matrix is
\be
\Pi = \begin{pmatrix}
	\frac{2 }{ \sqrt{3}} \,a\cdot \openone & H_1 & H_2 \\
	H_1^\dagger & (-\frac{1}{ \sqrt{3}}\,a + \sqrt{2}b)\cdot\openone & G \\
	H_2^\dagger & G^\dagger & (-\frac{1}{ \sqrt{3}}\,a -  \sqrt{2}b)\cdot\openone
	\end{pmatrix}
\ee
where  $H_{1,2}$ are electroweak doublets.  The fields $G$ are electroweak singlets which transform as a bi-doublet under  $SU(2)_{R1}\otimes SU(2)_{R2}$. 
Note that
\begin{align}
\sigma_2\, A - A^*\,\sigma_2 = 0 && \mathrm{where} \,\,A \equiv (H_i, G) \ .
\end{align}
The components of these fields  are parameterized by
\begin{align}
 H_i = \begin{pmatrix} 	h^i_0 + i h^i_3 &  i h_1  + h^i_2 \\
			i h^i_1 - h^i_2  & h^i_0 - i h^i_3
	\end{pmatrix}&& 
G = \begin{pmatrix}
	g_0 + i g_3 & i g_1 +  g_2 \\
	i g_1 - g_2 & g_0 - i g_3 
	\end{pmatrix} \ . 
\end{align}
These goldstone bosons transform as
\bea
H_1 &\to& V\,H_1\,U_1^\dagger \\
H_2 &\to& V\,H_2\,U_2^\dagger \\
G &\to& U_1\,G\,U_2^\dagger
\eea
where $V$ and $U_i$ are the $SU(2)_L$ and custodial $SU(2)_{Ri}$ representations, respectively. Note that as long as  only the neutral component of the Higgs fields  $H_{1,2}$ get a vacuum expectation value, and $\vev{G}=0$,  the three $SU(2)$s are broken to a diagonal $SU(2)_c$ for any ratio of the two vevs, and  the relationship, $m_w = m_z\,\cos{\theta_w}$ is preserved for any values of $\vev{H_{1,2}}$.  

 Defining $\theta = \vev{h^1_0}/ \sqrt{2} f$ and $\phi =\vev{ h^2_0}/\sqrt{2} f$, the sigma matrix becomes
\be
\Sigma = \begin{pmatrix}
	0	&	-i c	& 	0	& 	- c' s	& 	0 	& 	-s' s\\ 
	i c	&	0 	& 	-c'\, s	& 	0 	& 	 -s' s	& 	0 \\ 
	0	&  c' \,s & 	0	& 	-i (  { s'}^2+{c'}^2 c) 	& 	0 	& -i c' s'  (1-c) \\ 
	c' s & 	0 	& 	i (  { s'}^2+{c'}^2 c)	& 	0 	&  -i c' s' (1-c) & 0 \\ 
	0	&	 s' s	& 	0	& 	i c' s' (1-c)	& 	0 	& 	-i (  { c'}^2+{s'}^2 c)\\ 
	s' s &	0 	& -i c' s'(1-c)	& 	0 	& 	i (  { c'}^2+{s'}^2 c)	& 	0 	
	\end{pmatrix}
\ee
Here $s = \sin{(2 \sqrt{\theta^2 + \phi^2})}$, $c = \cos{(2\sqrt{\theta^2 + \phi^2})}$,  $c'=\theta/ \sqrt{\theta^2 + \phi^2}$,   and $s'=\phi/\sqrt{\theta^2 + \phi^2}$.
The tree level gauge boson masses are
\begin{align}
m_w^2 = 	\frac{g^2}{ 2}\,f^2 s^2 && m_z^2 = \frac{g^2 + g'^2 }{2}\, f^2 s^2\ .
\end{align}

As in the other models,
one-loop renormalization forces the introduction of the spurions
\be
-c g^2 f^4 \Tr Q \Sigma Q \Sigma^\dagger  -c g'^2 f^4 \Tr Y \Sigma Y \Sigma^\dagger + \mathrm{h.c.}
\ee
which, assuming positive $c$, are minimized when all  electroweak gauge bosons are massless.

\subsubsection{Yukawa Interactions}
As in the previous sections, we will induce the top mass by mixing the left and right handed components of the top with  new charge 2/3 quarks. The new quarks are members of multiplets $\Psi$, $\overline\Psi$ couple to the PNGBs in an $SU(6)$ preserving manner. We take $\Psi$, $\overline\Psi$ to be sextets under $SU(6)$.
\be
\begin{array}{cc}
\Psi \equiv \left( \begin{array}{ccccc}
 Q &  B_1 &  T_1 & B_2&  T_2\end{array} \right), & 
\overline{\Psi} \equiv \left( \begin{array}{c}
\overline{Q} \\ \overline{T}_1  \\  \overline{B}_1 \\ \overline{T}_2\\  \overline{B}_2 \end{array} \right) 
 \end{array}\ ,
\ee

The new quarks have the gauge quantum numbers
\begin{center}
\begin{tabular}{lcccc}
  &$SU(3)_c$& $SU(2)_L$ 	&  $U(1)_{Y}$	& $U(1)_Q$ \\ 
$Q $ &3	& 2			&	1/6		&  $\begin{pmatrix} -1/3 \\ 2/3 \end{pmatrix}$  \\ 
${T_{1,2}}$ 	&3	& 1				&  2/3 	& 2/3  \\
${B_{1,2}}$ &3		& 1		&  -1/3 	&  -1/3 \\
\end{tabular}
\end{center}
The Yukawa interaction of the top quark to the Higgs doublets arises from
\begin{equation}
  \label{topYukawa2}
  \CL_{t}=  \lambda_1 f \Psi \Sigma \overline{\Psi} + \lambda_2 f  \,{
  q_3} \overline{Q} + \lambda_3  f \, T_1 \overline{t}_3  + \lambda_4  f \, T_2 \overline{t}_3 + {\rm h.c.}
  \end{equation}
 A large top mass requires   $\lambda_{1,2}$ and either $\lambda_3$ or $\lambda_4$ to be of order 1.
The top coupling to $H_{1,2}$   requires nonvanishing  $\lambda_{3,4}$  respectively.  

\subsubsection{Additional terms}
As in the previous models, not all of the PNGBs receive mass from the gauge and Yukawa interactions. A small term in the effective theory proportional to
\be\label{mass} \Tr M \Sigma + h.c.\ee  
where $M$ is some  spurion matrix, can give mass to these bosons. In a UV completion where $\Sigma $ arises from a fermion condensate, this interaction could arise from a fermion mass term.
\subsubsection{Vacuum alignment}
Vacuum alignment in this model is complicated by the presence of two Higgs doublets. In a version where either $\lambda_3$ or $\lambda_4=0$, and with the additional term $M$ of  \ref{mass} chosen to be proportional to $\Sigma_0$, only one doublet will get a vev. 
In this simple case  the alignment analysis is nearly identical to that of the $SU(4)/SP(4)$ model. More generally
the alignment  and the angle $s'$  depend on all the $\lambda_i$'s and all parameters in the matrix $M$.  We will not attempt to  explore this rather large parameter space. 

 \section{Implications for Collider Signatures}
\label{sec: coll} 

In the intermediate Higgs scenario there are new vector-like quarks below a few TeV and  additional  spinless particles at the weak scale.  In the simple examples considered here the new scalars transform as electroweak singlets, doublets, and triplets. The leading terms in the potential for these scalars is generally computable in terms of a small number of symmetry breaking spurions,  gauge interactions, and Yukawa couplings. In most models there is a boson whose couplings to the $W$ and $Z$  are nearly those of the Higgs of the Minimal Standard Model.  Furthermore, in some models the Higgs mass may be computed in terms of the new fermion masses. Higgs search strategies are complicated by the new scalars, as the Higgs will generally have a substantial branching fraction for decay into the new scalars, which then decay into quarks. 

It is thus possible that the  Higgs  typically decays into four jets. Furthermore, in some models the dominant mechanism for scalar-quark interactions comes from mixing in the charge 2/3 sector, and so the dominant decays are into jets containing charge 2/3 quarks.  Such decays impact Higgs search strategies \cite{hep-ph/9908391,Dobrescu:2000jt,hep-ph/0411213,hep-ph/0502105}.  They also potentially weaken
the bound on the Higgs mass through Higgstrahlung processes off the $Z$ as seen by LEP.  This is especially true in regions of parameter space where the final decays of Higgs results into four well separated jets, which together with the $Z$ decay into 2 quarks is a six jet event.  These topologies have not been fully studied thus far, and may be consistent with a Higgs  mass below 114 GeV.  Note, to date, the most stringent analysis\footnote{We thank David E. Kaplan for discussion on this point.} on the impact of non-trivial decay modes on the Higgs' mass was done by the OPAL collaboration\cite{Abbiendi:2002qp}.  In the scenarios considered in this paper, the Higgs has a non-trivial decay fraction into a CP-even, neutral scalar.  With order one couplings between the two particle species and a small decay width, the lower bound on  Higgs' mass is $83.7$ GeV.

We have also seen that our choice of IH models, with a  top Yukawa sector that breaks custodial 
$SU(2)_c$ only softly, favors a  three spurion structure for the top Yukawa sector.  Generically, this sector  features additional charged $2/3$ quarks.  After going to the mass eigenstate basis, a linear combination of the charge 2/3 quarks becomes the top quark. 
In all the cases considered, there are at least four new quarks. Soft breaking of the custodial $SU(2)_c$ will generally require four or more additional quarks\footnote{To see this, note that the new quarks must be in a vector-like representations of the electroweak group, must be in a complete multiplet of custodial $SU(2)_c$, and must contain members with electroweak quantum numbers allowing mixing with the left and right handed top  in the limit of unHiggsed $SU(2)_L$.}.
In the case of $SU(5)/SO(5)$, we have  three new charge 2/3 quarks, a new charge -1/3 quarks, and a new charge 5/3 quark. The masses of all  five new quarks  as well as the mass of the top quark  are determined  by the four independent parameters $\lambda_{1,2,3}$ and $f$. In $SU(4)/SP(4)$, we have two new charge 2/3 quarks and two charge -1/3 quarks, and again all four masses and the mass of the top are determined by the four parameters $\lambda_{1,2,3}$ and $f$.
The case of $SU(6)/SP(6)$  involves six new quarks, three with charge 2/3 and three with charge -1/3. All six masses, together with the top mass, are determined in terms of  $\lambda_{1,2,3,4}, f$ and the ratio of the two Higgs vevs.  Thus, in principle,
 measurement of new quark masses  allows  verification of the symmetries and the structure of collective symmetry breaking
  without recourse to  extracting Yukawa couplings from  cross-sections as  in ref. 
\cite{Perelstein:2003wd}.   As an illustration, in the  limit of 
$v\ll f$, for $SU(5)/SO(5)$ this relation is simply 
\be
\label{massrel}
\lambda_t^2 = \frac{m_1^2(m_2^2-m_1^2)(m_3^2-m_1^2)}{f^2 m_2^2 m_3^2}
\ee
where $m_{1,2,3}$ are the masses of the three heavy charge 2/3 quarks, with $m_1$ being the lightest. The value of $f$ may be determined, {\it e.g.} from precise  measurement of the mass splittings of quarks which are electroweak doublets in the large $f$ limit. 

For $SU(4)/SP(4)$,  eq. \ref{massrel} also holds in the $v\ll f$ limit, with $m_1$ being the mass of the lighter of the new charge -1/3 quarks, and $m_{2,3}$ being the masses of the two new charge 2/3 quarks. 
The ratio $v/f$ can be determined in principle from the splitting between the mass of the heavier of the charge -1/3 quarks and its electroweak partner charge 2/3 quark. 

\section{Experimental Constraints}
\label{sec: exp}

\subsection{Precision Electroweak Constraints}
\label{sec:pewc}

Our low energy effective theory is relatively economical in its new particle content and so has few possible sources of precision electroweak corrections. The low energy effective theory possesses an approximate custodial $SU(2)_c$ symmetry, broken mainly by the hypercharge and top Yukawa couplings, which eliminates  large  tree level corrections to the $T$ parameter. There will be one loop corrections to the $T$ parameter from the top sector, which are altered relative to the one loop corrections of the Standard Model by a factor of order $m_t^2/m_{T'}^2$, where $m_{T'}$ is the mass scale of the new heavy quarks.  Similarly, the one loop corrections to the coupling of the left handed $b$ quark to the Z boson can  be altered  by a factor of this order. Details of these corrections  are the same as those of any theory with new quarks, as discussed, e.g.,  in refs.~\cite{Cohen:1983fj, Lavoura:1992np,Collins:1999rz,Popovic:2000dx,Popovic:2001cj,He:2001fz}.
In addition, there is the familiar operator contributing to the $S$ parameter in technicolor theories, $1/(4\pi)^2 \tr(W_{\mu\nu}\Sigma B^{\mu\nu}\Sigma^\dagger)$, but as mentioned, this is 
suppressed by $(v/f)^2$. A rough estimate on the bounds on $v/f$ may be obtained by noting that technicolor theories typically have corrections to $S$ and $T$ which are estimated to be too large by a factor of three or four, so $v/f \lesssim 1/2$ should be roughly sufficient to ensure agreement with low energy experiment.

\subsection{Flavor Changing Neutral Currents} 

The cutoff scale of these models is low enough that Flavor Changing Neutral Currents (FCNC) will severely constrain the UV completion, as in essentially all models of physics beyond the Standard Model.  Acceptable UV completions for FCNC include high scale (50 TeV) supersymmetry \cite{Katz:2003sn}, a warped extra dimension \cite{hep-ph/0412089,hep-ph/0501036}, or, equivalently  a strongly coupled nearly conformal theory. 
Within the low energy effective theory, the additional vector-like quarks can mix with the light quarks and give new sources of  FCNC. In ref. \cite{hep-ph/0408362} it was shown that  a natural expectation is that the heavy charge 2/3 states will mix slightly with the light charge 2/3 quarks and produce FCNC in the charge 2/3 sector which are  much larger than those of the minimal Standard Model, although well below the experimental bounds. For instance, the decay $t\rightarrow cZ$ is expected to have a branching fraction which is much larger than in the Standard Model, although probably too small to be observed at the LHC.  A more general study of the contribution of the heavy quarks  to $b$ and $c$ physics in these models would be very interesting.

\section{UV completions of the Intermediate Higgs}
 \label{sec: UV} 
 
In all these models the Higgs and its fellow PNGBs become strongly coupled  at energies above   $4 \pi f$.  We take the cutoff of the effective theory to be at or below this scale. The symmetry breaking pattern $G/H$ could be produced by the dynamics of a new strong interaction, in a manner analogous to the breaking of chiral symmetry by QCD. Like the QCD pions, the Higgs boson would then be a composite particle.
$SU(n)$ is a typical approximate symmetry   whenever there are $n$ fields in the same representation of some strong  gauge group. The breaking pattern $SU(n)/SO(n)$ results from the symmetric condensate of a real representation while $SU(n)/SP(n)$ is produced by the antisymmetric condensate of a pseudoreal representation\footnote{Note that Thaler \cite{hep-ph/0502175} has shown how to spontaneously break a global $SU(N)$ symmetry  to various subgroups $H$ using  two new gauge groups---one with strong QCD-like dynamics and another which is moderately weak.}.

A mechanism to produce the Higgs Yukawa  couplings, involving composite $T'$'s as well as a composite Higgs,    has been discussed in ref. \cite{Katz:2003sn}. In that work, heavy scalars stabilized by high scale supersymmetry produced certain required four fermion couplings.  It is also possible that, as  discussed in ref. \cite{Katz:2003sn,hep-ph/0409274}, 
if the UV theory is strongly coupled and nearly conformal, the necessary four fermion couplings might acquire a large  anomalous dimension and be nearly marginal. While low energy precision physics in  such a picture is problematic  for technicolor theories, in composite intermediate or little Higgs scenarios the precision electroweak corrections are reduced by $(v/f)^2$ and could be acceptable.

Alternatively,  our low energy effective theory could arise from a Randall-Sundrum type  model \cite{hep-ph/9905221} with a warped 5th dimension with gauge group $G$ in the bulk broken to $H$ by the boundary conditions on an IR brane, as in  refs. 
\cite{hep-ph/0412089,hep-ph/0501036}.  Such a model  presumably has a dual description in terms of a four dimensional theory with a new strongly coupled nearly conformal  gauge group \cite{hep-th/9711200,hep-th/0012148,hep-th/0012248}.

 \section{Summary and conclusions}
\label{sec: con}

In the next decade,
 colliders will explore the multi TeV range, where we hope that new particles to be discovered which will elucidate the origin of electroweak superconductivity. In this paper we propose a new approach to  modeling the Higgs as a pseudo Nambu-Goldstone boson. We restrict ourselves to an effective field theory with a range of validity up to  about 6 TeV. With such a low cutoff, the finetuning problem of the Standard Model arises only from the one loop UV sensitivity of the Higgs potential   from the top Yukawa coupling. We therefore  consider models in which the top  mass arises from  the collective symmetry breaking mechanism.  We explore  several minimal models. These all have distinctive low energy phenomenology, with additional scalars at the electroweak scale, and new  vector-like  quarks. The new quarks may be searched for in hadron colliders, and have  interesting implications  for low energy flavor physics. The nonminimal Higgs sector will impact the strategy for Higgs searches.
 
 \section{Acknowledgements}

The work of E. Katz is supported by DOE grant DE-FG02-91ER40676 and DE-AC02-76SF0051.  A. Nelson was partially supported by the DOE under contract DE-FGO3-96-ER40956.  D. Walker is supported by a NSF grant NSF-PHY-9802709 and a grant from the Ford Foundation.  We would like to thank Nima Arkani-Hamed, Spencer Chang, Hsin-Chia Cheng, David B. Kaplan, David E. Kaplan, Ian Low and Markus Luty for useful discussions and critical revisions of the draft.

\end{document}